\documentclass[12pt]{article}
\usepackage{amsmath,epsf,amssymb,latexsym,cite}
\newtheorem{prop}{Proposition}

\newif\iffigs\figstrue

\DeclareFontFamily{U}{rsf}{}
\DeclareFontShape{U}{rsf}{m}{n}{
  <5> <6> rsfs5 <7> <8> <9> rsfs7 <10-> rsfs10}{}
\DeclareMathAlphabet\Scr{U}{rsf}{m}{n}

\def\pplogo{\vbox{\kern-\headheight\kern -29pt
\halign{##&##\hfil\cr&{
\ppnumber}\cr\rule{0pt}{2.5ex}&\ppdate\cr}
}}
\makeatletter
\def\ps@firstpage{\ps@empty \def\@oddhead{\hss\pplogo}%
  \let\@evenhead\@oddhead 
}
\def\maketitle{\par
 \begingroup
 \def\thefootnote{\fnsymbol{footnote}}
 \def\@makefnmark{\hbox{$^{\@thefnmark}$\hss}}
 \if@twocolumn
 \twocolumn[\@maketitle]
 \else \newpage
 \global\@topnum\z@ \@maketitle \fi\thispagestyle{firstpage}\@thanks
 \endgroup
 \setcounter{footnote}{0}
 \let\maketitle\relax
 \let\@maketitle\relax
 \gdef\@thanks{}\gdef\@author{}\gdef\@title{}\let\thanks\relax}
\makeatother

\def\O{{\mathcal O}}

\def\P{{\mathbb P}}

\def\R{{\mathbb R}}
\def\Z{{\mathbb Z}}

\def\Pic{\operatorname{Pic}}

\def\Grm{\operatorname{Grm}}
\def\Gr{\operatorname{Gr}}

\def\Sl{\operatorname{SL}}
\def\GO{\operatorname{O{}}}
\def\SU{\operatorname{SU}}
\def\GU{\operatorname{U{}}}

\def\Spin{\operatorname{Spin}}

\def\CY{Calabi--Yau}

\def\cM{{\Scr M}}

\def\cG{{\Scr G}}
\def\ff#1#2{{\textstyle\frac{#1}{#2}}}

\def\spnh{\Spin(32)/\Z_2}

\begin{document}
\setcounter{page}0
\def\ppnumber{\vbox{\baselineskip14pt\hbox{DUKE-CGTP-99-08}
\hbox{hep-th/9910248}}}
\def\ppdate{October 1999} \date{}

\title{\LARGE Heterotic String Corrections \\
  from the Dual Type II String\\[10mm]}
\author{
Paul S. Aspinwall and M. Ronen Plesser\\[10mm]
\normalsize Center for Geometry and Theoretical Physics, \\
\normalsize Box 90318, \\
\normalsize Duke University, \\
\normalsize Durham, NC 27708-0318\\[10mm]
}

{\hfuzz=10cm\maketitle}

\def\Large{\large}
\def\LARGE{\large\bf}


\begin{abstract}
We introduce a method of using the a dual type IIA string to compute
$\alpha'$-corrections to the moduli space of heterotic string
compactifications. In particular we study the hypermultiplet moduli
space of a heterotic string on a K3 surface.  One application of this
machinery shows that type IIB strings compactified on a \CY\ space
suffer from worldsheet instantons, spacetime instantons and, in
addition, ``mixed'' instantons which in a sense are both worldsheet
and spacetime.  As another application we look at the hyperk\"ahler
limit of the moduli space in which the K3 surface becomes an ALE
space. This is a variant of the ``geometric engineering'' method used for
vector multiplet moduli space and should be applicable to a wide range
of examples. In particular we reproduce Sen and Witten's result for
the heterotic string on an $A_1$ singularity and a trivial bundle and
generalize this to a collection of $E_8$ point-like instantons on an
ALE space.
\end{abstract}

\vfil\break


\section{Introduction}    \label{s:int}

At the present time theories with eight supercharges (such as $N=2$ in
four dimensions) represent the ``Goldilocks'' theories for analyzing
moduli spaces of superstring compactifications. 
With more supersymmetry the moduli space is a symmetric space and in
many respects is too rigid to be particularly interesting. With less
supersymmetry quantum effects lift some of the classical moduli and
computing these is in general not possible with current techniques.

For $N=2$ theories in four dimensions, the moduli space factorizes
locally into a product $\cM_V\times\cM_H$, where $\cM_V$ is the
special K\"ahler moduli space coming from scalar fields in vector
multiplets and $\cM_H$ is the hypermultiplet moduli space which is
quaternionic K\"ahler. The space $\cM_V$ has been studied at great
length and can be considered to be fairly well understood. The same is
not true for $\cM_H$.

An important obstacle in the study of $\cM_H$ can be understood as
follows. If we wanted to construct a model producing $N=2$
supersymmetry (complete with gravity) in four dimensions we might try
to do it in one of many ways. For example:
\begin{enumerate}
\item Compactify the type IIA string on a \CY\ threefold $X$.
\item Compactify the type IIB string on a \CY\ threefold $Y$.
\item Compactify the heterotic string on $S_H\times E_H$, where $S_H$
is a K3 surface and $E_H$ is a 2-torus.
\end{enumerate}

In each of these pictures some aspects of the moduli space are {\it
exact}, in the sense that a computation at string and $\sigma$-model
tree level is not corrected by quantum corrections, 
and some aspects are prone to $\alpha'$-corrections, i.e., quantum
corrections for the nonlinear $\sigma$-model; or $\lambda$-corrections,
i.e., quantum corrections from the effective spacetime quantum field
theory (where $\lambda$ represents the string coupling). We list these
effects in table~\ref{tab:qc}.

What is important to note is that the type IIB string on $Y$ gives an
{\em exact\/} model for the moduli space $\cM_V$ whereas none of these
pictures gives an exact model for $\cM_H$. Indeed we do not know any
way to get an exact picture of $\cM_H$. If we could map out
$\cM_H$ then this would be the first time in our study of string theories
that we were able to probe truly nonperturbative effects both with
respect to $\alpha'$ and $\lambda$. It could well be therefore that a
better understanding of $\cM_H$ would lead to a better understanding
of the fundamental nature of string theory in general.

What we can see about table~\ref{tab:qc} is that the compactifications
listed above tend to shine light on each other's ignorance about
$\cM_H$. In particular one could try to use the type IIA string to
understand the $\alpha'$-corrections in the heterotic string
compactification. 

The general problem of completely analyzing $\cM_H$ complete with all
its quantum corrections is technically very difficult at present
although there appears to be nothing lacking in our knowledge in
principle. In this paper we will attempt a rather modest goal. 
For the latter part we will confine ourselves to the hyperk\"ahler limit
of the moduli space where most of the interesting string corrections disappear.
We will rederive a result already computed in two different ways by Sen
\cite{Sen:kkm} and by Witten \cite{W:hADE} and then try to generalize this
result. We show that the moduli space of heterotic
strings on a blown-up $A_1$ singularity, with a trivial bundle near
this singularity, is locally described by a hyperk\"ahler limiting
manifold of real dimension 4 given by an Atiyah--Hitchin monopole moduli
space. Equivalently the moduli space is the same as that of a pure
$N=4$ $\SU(2)$ Yang--Mills theory in three dimensions \cite{SW:3d}.

\begin{table}
\begin{center}
\begin{tabular}{|c||c|c|}
\hline
&$\cM_H$-corrections&$\cM_V$-corrections\\
\hline
IIA on $X$&$\lambda$&$\alpha'$\\
IIB on $Y$&$\lambda$ and $\alpha'$&Exact\\
Het on $S_H\times E_H$&$\alpha'$&$\lambda$\\
\hline
\end{tabular}
\end{center}
\caption{Quantum corrections.}
        \label{tab:qc}
\end{table}

We would like to claim that our approach to this problem is useful in
many ways. Firstly it is an example of using duality to the type IIA
string to obtain nontrivial results about $\alpha'$-corrections to the
heterotic string. Secondly, it provides a natural link between the
heterotic strings on an ADE singularity and the moduli space of 
corresponding $N=4$ Yang--Mills theories in three dimensions. 

On a third point our method shows that taking the ``rigid limit'' in
which we move from quaternionic K\"ahler geometry to hyperk\"ahler
geometry is mirror to taking the rigid limit of special K\"ahler
geometry in vector multiplet moduli space to obtain Seiberg--Witten
theory. Thus the ``geometric engineering'' methods of \cite{KKL:limit,
KKV:geng} of analyzing Seiberg--Witten theories can be modified using
mirror symmetry to ``geometrically engineer'' statements about the
the hyperk\"ahler moduli space of hypermultiplets in a corresponding limit.
This idea of using mirror symmetry together with geometric engineering
to obtain a hypermultiplet moduli space plays a role in various
contructions that have appeared in the past
\cite{GMV:Ht,OV:Din,SS:3dU,KMV:l-mir1}. In the context of using it to
describe vector bundles in the heterotic string it also appears
to be closely related to the construction presented in \cite{BM:Fbun}.

One should be able to generalize the computation in this
paper to many statements about the heterotic string with various
bundles on ALE spaces. As an example we ``engineer'' the problem of
one or two free point-like $E_8$ instanton coalescing with an $A_1$
singularity. This generalizes to give a natural proposal:
\begin{prop}
The moduli space of a heterotic string on an ALE space of type
$A_{n-1}$ with $k$ point-like $E_8$ instantons is same as the moduli
space of an $N=4$ gauge theory in three dimensions with gauge group
$\SU(n)\times\GU(1)^k$ and $k$ matter hypermultiplets each in the
fundamental representation of $\SU(n)$ and charged with respect to one
of the $\GU(1)$'s.\footnote{To be pedantic this group should probably
be written $\SU(n)\times\GU(1)^k\rtimes\mathsf{H}$, where 
$\mathsf{H}$ is a discrete group containing as a subgroup the 
symmetric group on $k$ elements. We will ignore
such discrete factors here.} \label{p:d1}
\end{prop}

In principle our method can be used to extend this result to any
kind of bundles, including smooth bundles, point-like $\spnh$
instantons or fractional point-like instantons as in \cite{AM:frac}. 

It should be emphasized however that this paper should be considered
to be the first step in addressing the more interesting question of the
full quaternionic K\"ahler moduli space. In a sense by going to the
hyperk\"ahler limit we are avoiding the truly ``stringy'' nature of
the subject and confining ourselves to statements about quantum field
theory.

In order to simplify the exposition, this paper will be oriented
towards studying a particular example 
although we will be able to make general statements based on our
analysis. It would be nice to study an example where the
dimension of $\cM_H$ is fairly small. This in itself is not
particularly easy since simple compactifications of the heterotic
string on $S_H\times E_H$ tend to produce rather enormous $\cM_H$'s as
seen in \cite{KV:N=2} for example. Our example will have a moduli
space of only 4 quaternionic dimensions.

The general principle of the manipulations in this papers are as
follows. First we find a \CY\ threefold $X$ which has a stable
degeneration such that the IIA string on $X$ is, at this point, dual to
a weakly coupled, large radius limit of 
the heterotic string configuration we
desire. If $Y$ is the mirror of $X$ we will find that, at least in the
class of examples we consider, $Y$ is a K3 fibration with a $\P^1$
base. As we move to the stable degeneration of $X$, the base of $Y$
becomes infinitely large. As such the stable degeneration of $X$ is
mirror to to the way that Seiberg--Witten theories were
``geometrically engineered'' in \cite{KKL:limit} from the type IIA
string compactified on $Y$. We will engineer hyperk\"ahler moduli
spaces as rigid limits of the quaternionic K\"ahler manifold $\cM_H$
rather than the original process of engineering rigid special K\"ahler
moduli spaces as rigid limits of the special K\"ahler manifold $\cM_V$.

We introduce the example in section \ref{s:lim1} and we compute the
stable degeneration required to understand the duality to the
heterotic string. We map out the moduli space of this degeneration in
section \ref{s:cl}.

In section \ref{s:close} we discuss what happens if we try to move
away from the stable degeneration back into the bulk of the moduli
space. The central point of this section is that the K3 surface $S_H$ of
interest is an elliptic fibration with an infinitely large section and
a fibre of zero area. Its volume is finite however and the generic
areas of the transcendental 2-spheres are also finite and nonzero.

The picture obtained in section \ref{s:close} will allow us to make
some brief comments about ``mixed instantons'' in section
\ref{s:IIB}. We argue that there are instanton corrections beyond what
one would na\"\i vely consider necessary in superstring compactifications.

In section \ref{s:hypK} we then discuss another application of section
\ref{s:close}. Essentially this is analysis of the ``first-order''
effects as we move back into the bulk of the moduli space. This
studies a hyperk\"ahler limit of the general quaternionic K\"ahler
moduli space. In particular this allows us to recover Sen and Witten's
result. It is then a simple matter to introduce point-like $E_8$
instantons into the picture and derive proposition \ref{p:d1}. We then
show how this proposition is linked to Intriligator--Seiberg mirror
symmetry in three dimensions.

While we were putting the final touches to this manuscript,
\cite{Roz:hyp3d} appeared which derives proposition~\ref{p:d1} by
using M-theory arguments.


\section{An Example} \label{s:lim1}

Let us begin with the heterotic/type IIA pair introduced in
\cite{KV:N=2} and later clarified in \cite{MV:F2}. We consider $Y$
to be the hypersurface
\begin{equation}
  x_0^2+x_1^3+x_2^{12}+x_3^{24}+x_4^{24}=0 \label{eq:eqnX0}
\end{equation}
in the weighted projective space $\P^4_{\{12,8,2,1,1\}}$. The type IIA
string on $Y$ is then believed to be dual to an $E_8\times E_8$
heterotic string compactified on $S_H\times E_H$ where the gauge
bundle is trivial over $E_H$ while over $S_H$ we have two $E_8$ bundles
with $c_2=10$ and $c_2=14$ respectively.

The moduli space $\cM_H$ is then constructed in the heterotic language
from deformations of $S_H$ together with its bundle. A K3 surface has
20 quaternionic deformations and an $E_8$ bundle has $30c_2-248$ 
quaternionic deformations. Thus the quaternionic dimension of $\cM_H$
in this case would be 244. Alternatively from the type IIA side
$\cM_H$ comes from deformations of complex structure of $Y$,
variations of the Ramond-Ramond fields in $H^3(Y,\GU(1))$ and the
dilaton-axion. This gives $h^{2,1}(Y)+1=244$ quaternionic dimensions
again.

Clearly it would be a very unwieldy to study an example with this many
deformations! In order to cut down to number of hypermultiplets we
need to tighten our constraints on the $E_8$ bundles and tighten our
constraints on the K3 surface $S_H$ itself.

Note that this example has $h^{1,1}(Y)=3$. We would be much better off
if we took the mirror of $Y$, which we call $X$, and considered the
type IIA string on $X$. Now we would have only 4 hypermultiplets.
The mirror pair $X$ and $Y$ have been studied extensively
\cite{KLM:K3f,LY:K3f}. The discussion in \cite{VW:pairs} is also useful
for our purposes.

As well as the usual Greene--Plesser orbifold \cite{GP:orb}
construction we may also write $X$ in the form of a singular version
of $Y$. Let $X$ be the hypersurface in $\P^4_{\{12,8,2,1,1\}}$ given
by the defining equation
\begin{multline}
  f_X=a_0x_0x_1x_2x_3x_4 + a_1x_2^6x_3^6x_4^6 + a_2x_2^5x_3^7x_4^7 +\\
  a_3x_2^7x_3^5x_4^5 + a_4x_2^5x_3^6x_4^8 + a_5x_2^5x_3^8x_4^6 +
  a_6x_1^3 + a_7x_0^2=0.   \label{eq:eqnX}
\end{multline} 
We implicitly
assume that $X$ is blown up to render it smooth. This exhibits an
extremal transition from $Y$ to $X$. We now have 3 
deformations of complex structure which may be conveniently
represented by
\begin{equation}
\begin{split}
  x &= \frac{a_0^6}{a_1a_6^2a_7^3}\\
  y &= \frac{a_4a_5}{a_2^2}\\
  z &= \frac{a_2a_3}{a_1^2}.
\end{split}   \label{eq:xyz}
\end{equation}

Now the stable degeneration method of \cite{FMW:F,AM:po} may be
used to recover exactly which $E_8\times E_8$ heterotic string this
type IIA compactification represents. In particular we need to find a
form of $X$ as a K3 fibration and then find a degeneration which takes
each K3 fibre to a union of two rational elliptic surfaces
intersecting along an elliptic curve. It is this process which gives a
picture of a weakly-coupled heterotic string compactified on a large
K3 surface allowing us to make an unambiguous identification of the
heterotic string compactification.

\iffigs
\begin{figure}[t]
\begin{center}
  \epsfysize=11cm\leavevmode\epsfbox{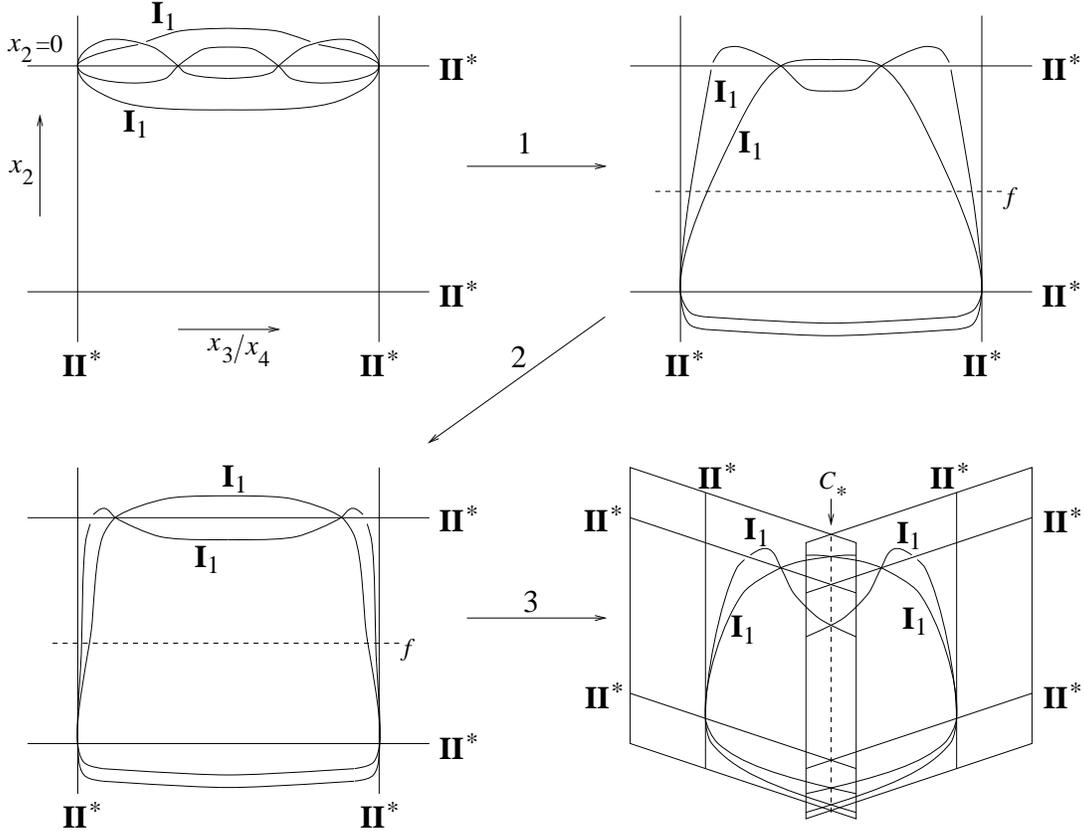}
\end{center}
  \caption{The stable degeneration of $X$.}
  \label{fig:squig}
\end{figure}
\fi

This process is not particularly direct. We depict the manipulations
involved in figure~\ref{fig:squig}. The pictures are those typical in
F-theory. They show the \CY\ threefold in the form of an elliptic
fibration where the plane of the paper represents the base and the
lines represent the discriminant locus labelled by the corresponding
Kodaira fibre. For more information we refer to \cite{me:lK3}.

\def\HS#1{{\bf F}_{#1}}
The process is as follows:
\begin{enumerate}
\item We begin with $X$. This has four lines of $\mathrm{II}^*$ fibres
as shown in the upper-left figure. The base is the Hirzebruch surface
$\HS2$. We 
wish to change the base to $\HS0\cong\P^1\times\P^1$. This may be done
as explained in \cite{MV:F2} by blowing up and down curves in the
base. In particular we blow up the two top corners of the square in
the top-left figure and then blow down the lines corresponding to the
left and right sides of the square. This yields the top-right
figure. Note that this birational transformation does not change the
moduli space of complex structures of $X$.
\item Now we begin to deform the complex structure by taking
$y\to0$. The interesting parts of the discriminant move towards the
vertical sides of the square. If we restrict the elliptic fibration to
a horizontal line $f$ we obtain an elliptic K3 surface with two
$\mathrm{II}^*$ fibres and four $\mathrm{I}_1$ fibres. As we take
$y\to0$ the $\mathrm{I}_1$'s move pairwise into the $\mathrm{II}^*$
fibres. This is exactly the stable degeneration required as seen in
\cite{FMW:F}.
\item When $y=0$ we obtain very serious singularities along the
vertical sides of the square. We may improve matters by blowing up up
these sides to produce two new components of the base. Furthermore the
original component of the base may be blown down without loss of
information. We end up with a base with two components as expected for
the stable degeneration picture. The components intersect along a
rational curve $C_*$. Restricting the elliptic fibration to $C_*$
gives a K3 surface $S_*$. This surface $S_*$ is then isomorphic to the
K3 surface $S_H$ 
on which the $E_8\times E_8$ heterotic string is compactified.
\end{enumerate}

The standard F-theory rules then tell us that $S_H$ is a K3 surface
with two $E_8$ quotient singularities (as two $\mathrm{II}^*$ fibres
pass through $C_*$). Furthermore, the rules in \cite{AM:po} tell us
that all 24 instantons are point-like and 10 of them sit at one of the
singularities and 14 sit at the other singularity.

This shows explicitly why we have only 4 hypermultiplets. 16 of the 20
deformations of the K3 surface have been fixed by demanding two $E_8$
singularities. All deformations of the bundle have been fixed by
stating that the instantons are point-like and that their locations
are fixed. We assume that $X$ has been blown up, i.e., the
corresponding vector or tensor moduli have been switched on to move us
into the ``Coulomb phase''  thus freezing all other hypermultiplet moduli.

It is perhaps worth noting that this stable degeneration and resulting
heterotic interpretation of $X$ is not standard. In the usual F-theory
language of Morrison and Vafa \cite{MV:F,MV:F2} the K3 fibration of
$X$ is given by restricting the elliptic fibration to vertical lines
in figure~\ref{fig:squig}. We have restricted the elliptic fibration
to horizontal lines $f$. Our approach is therefore related to the
conventional picture by heterotic-heterotic duality as discussed in
\cite{DMW:hh,AG:mulK3,MV:F} where the vertical and horizontal
directions are exchanged.
This appears to be an essential step if we want the stable
degeneration of $X$ to correspond to the mirror of giving the base
$\P^1$ of $Y$ a very large area.


\section{Mapping out the Completely Classical Limit}  \label{s:cl}

To recap, we are going to study an example where the number of
quaternionic moduli for $\cM_H$ is equal to four. This has three
interpretations:
\begin{enumerate}
\item A type IIA string (or F-theory) compactified on $X$. $X$ has
three deformations of complex structure as given by
(\ref{eq:eqnX}). The other moduli giving $\cM_H$ are 8 real R-R moduli
and the dilaton-axion.
\item A type IIB string on $Y$, the mirror of $X$. $Y$ has
three deformations of K\"ahler form given by the size of the original 
$\P^4_{\{12,8,2,1,1\}}$ together with two blow-ups required to smooth
$Y$. The other moduli giving $\cM_H$ are 8 real R-R moduli
and the dilaton-axion.
\item A heterotic string on a K3 surface $S_H$ (times a
2-torus). $S_H$ has two $E_8$ singularities leaving only 4
quaternionic deformations of the Ricci-flat metric and $B$-field. The
bundle is completely rigid and corresponds to point-like instantons
buried in the singular points of $S_H$.
\end{enumerate}

Now {\em none\/} of these three pictures gives an exact model for the
entire moduli space. The type IIA picture will give an exact picture if we
fix the dilaton so that the string coupling is zero. This is the limit
we are interested in for this paper.

Let us introduce the notation
\begin{equation}
\Grm(\Lambda)=\GO(\Lambda)\backslash\GO(n,m)/(\GO(n)\times\GO(m)),
\end{equation}
where $\Lambda$ is a Lorentzian lattice (which may or may not be even and
unimodular) which has signature $(n,m)$. Thus the moduli space of type
IIA strings on a K3 surface would be written $\Grm(\Gamma_{4,20})$,
where $\Gamma_{m,n}$ will denote an even unimodular lattice of
signature $(n,m)$.

In our case we decompose the lattice $H^*(S_H,\Z)$ into $\Gamma_{4,20}
\cong\Gamma_{4,4}\oplus\Gamma_{0,8}\oplus\Gamma_{0,8}$ where the
latter two lattices represent time-like root lattices of
$E_8$. In order to fix the two $E_8$ singularities and set the
corresponding components of the $B$-field equal to zero we should
delete these latter parts of the $H^*(S_H,\Z)$ lattice as explained in
\cite{me:lK3} for example. Thus, {\em if there were no
$\alpha'$-corrections\/} we might expect
\begin{equation}
  \cM_H = \Grm(\Gamma_{4,4})  \label{eq:44}
\end{equation}
from the heterotic picture.

To make contact with the type IIA picture we follow the usual
F-theory process as proposed in \cite{FMW:F,AM:po} (see also
\cite{me:lK3}). Let us suppose that $S_H$ is actually an elliptic
fibration with a section. We may do this as follows. The
$\Gamma_{4,4}$ in (\ref{eq:44}) represents $H^*(S_H,\Z)$ without the
$E_8\oplus E_8$ part coming from the singularities. Let us split this
as $\Gamma_{4,4}=\Gamma_{2,2}^A\oplus\Gamma_{2,2}^T$. Let us split
further $\Gamma_{2,2}^A=\Pic(S_H)\oplus H^0(S_H,\Z)\oplus
H^4(S_H,\Z)$, where $\Pic(S_H)=\Gamma_{1,1}$ is the Picard lattice of
a generic elliptic K3 surface with a section. Note that $\Gamma_{2,2}^T$
represents the ``transcendental'' 2-cycles of $S_H$. These 2-cycles
have homology classes which are orthogonal to those of the algebraic curves.
Any homology 2-cycle can be split into its algebraic part in
$\Pic(S_H)$ and its transcendental part in $\Gamma_{2,2}^T$.

We now have a natural embedding
\begin{equation}
  \Grm(\Gamma_{4,4}) \supset \Grm(\Gamma_{2,2}^A)\times\Grm(\Gamma_{2,2}^T).
\end{equation}
$\Grm(\Gamma_{2,2}^A)$ is a 4 real-dimensional space spanned by
the value of the complexified K\"ahler form $B+iJ$ on the fibre and
the section of $S_H$ respectively. $\Grm(\Gamma_{2,2}^T)$ is a 4
real-dimensional space representing the remaining deformations of
complex structures of $S_H$ as an elliptic fibration. Note that
$\Grm(\Gamma_{4,4})$ had 16 real dimensions and so we have lost 8 real
dimensions in the process of insisting that $S_H$ be an elliptic
fibration with a section.

Now start taking limits. First we want to take the type IIA
string to be weakly coupled. The dilaton of the type IIA string
corresponds to the size of the section of $S_H$ \cite{me:lK3}. Thus we
may take the type IIA string to be weakly coupled by moving in the
right direction in $\Grm(\Gamma_{2,2}^A)$ so as to make the area of
the section huge.

If we take the other component of the K\"ahler form to be huge so that
the area of the elliptic fibres of $S_H$ are also large area then we
make every 2-cycle in $S_H$ huge (assuming the complex structure is
generic). This will remove all $\alpha'$ corrections. This process will
involve moving in a certain direction in the moduli space of complex
structures of the threefold $X$ which results in a stable
degeneration of $X$. This stable degeneration is of course the one
described in section \ref{s:lim1}. In this
limit we expect the remaining two deformations of complex structure of
$X$ to map {\em exactly\/} onto $\Grm(\Gamma_{2,2}^T)$.

Fortunately all the hard work of performing the identification in
this example has already been done elsewhere \cite{KLM:K3f,LY:K3f,KKL:limit}.
Let us recall the variables defined by (\ref{eq:xyz}).
The required stable degeneration is then given by the limit
$y\to0$. In this limit there is an explicit map between $x$ and $z$
and $\Grm(\Gamma_{2,2}^T)$ given as follows. First note that there is
an isomorphism between $\Grm(\Gamma_{2,2}^T)$ and two $j$-lines up to
discrete identifications:
\begin{equation}
\Grm(\Gamma_{2,2}^T)\cong(\Z_2\times\Z_2)\backslash\left(
  \frac{\Sl(2,\R)}{\GU(1)}\times\frac{\Sl(2,\R)}{\GU(1)}\right). \label{eq:jj}
\end{equation}
We refer to \cite{Giv:rep} for example for details of this isomorphism.
Then we may map the variables parametrizing the moduli space of $X$ in
(\ref{eq:xyz}) to two copies of the $j$-line by letting $j_1$ and
$j_2$ be the two solutions of
\begin{equation}
  zj^2+(432-x-1728z)j+x^2=0.  \label{eq:jj1}
\end{equation}

All this looks very similar to previous work done in the vector moduli
space such as in Kachru et al.~\cite{KKL:limit}. Indeed we will follow
this work further but it is important to remember that we are working
in the hypermultiplet moduli space and not the vector multiplet space. In
the vector multiplet moduli space case studied in \cite{KKL:limit} the
moduli space given by (\ref{eq:jj}) is the special K\"ahler moduli
space coming from the {\em torus\/} for the heterotic string. In our
case this moduli space (\ref{eq:jj}) is the moduli space of complex
structures on the {\em K3 surface\/} for the heterotic string.  In a
sense we are working in the mirror side with respect to
\cite{KV:N=2,KKL:limit}. We consider the type IIA string on $X$
whereas \cite{KV:N=2,KKL:limit} consider the type IIB string on the
same $X$.

We began with a moduli space of 4 quaternionic dimensions and we have
ended up with a 4 real dimensional space in this stable degeneration limit.
Let us review the moduli we have fixed or lost by going to this
limit. 
For the type IIA string on $X$:
\begin{enumerate}
\item We have lost the 8 R-R moduli which have no effect in the zero
string coupling limit.
\item In the type IIA picture we have fixed the dilaton to zero string
coupling and therefore lost the axion which now lives on a circle of
zero radius. 
\item We have fixed one complex deformation to go the the stable
degeneration limit.
\end{enumerate}
Correspondingly in the heterotic string picture:
\begin{enumerate}
\item We have fixed the K3 surface $S_H$ to be elliptic with a
section. This fixes 8 real moduli.
\item We have taken the section to be infinite in area. This kills the
effect of the associated $B$-field.
\item We have taken the elliptic fibre to have infinite area. This kills the
effect of the associated $B$-field.
\end{enumerate}
In both cases we account for the 12 missing moduli.


\section{Close to the Limit}   \label{s:close}

Studying a stable degeneration in this case doesn't tell us anything new except
for the fact that heterotic/type IIA duality seems to work nicely. It
is more interesting if we try to move away from this limit.

In particular what happens if we keep the type IIA string coupling
constant zero (or at least very small) but we move away from the
stable degeneration? The type IIA picture will remain exact since we
aren't switching on any $\lambda$-corrections.
On the heterotic side however we appear to be allowing for some
$\alpha'$-corrections to appear.

At first sight you might think that, in terms of the heterotic string,
we should study the part of the moduli space where the section of
$S_H$ is infinite and we give some nonzero value to the size of the
fibre. This is not the case and we need to study this process with a
little more care. 

As Witten argued in \cite{W:K3inst}, any $\alpha'$-corrections to the
moduli space of the heterotic string should come from world-sheet
instantons wrapping themselves around minimal 2-spheres in $S_H$. In the
stable degeneration of section \ref{s:lim1} we have an elliptic K3
surface with infinitely large section and fibre. Assuming the complex
structure is generic this would imply that {\em every\/} 2-sphere in
$S_H$ has infinite area and so there are no quantum corrections.

Let us consider more carefully the finite case away from this
limit. We will describe the moduli space $\Grm(\Gamma_{4,4})$ in terms
introduced in \cite{AM:K3p,me:lK3}.
Let $\Gamma_{4,4}=\Gamma_{1,1}\oplus\Gamma_{3,3}$ where $\Gamma_{1,1}$
is spanned by a light-like vector $w$ and its dual $w^*$. A point in 
$\Grm(\Gamma_{4,4})$ is determined by a space-like 4-plane 
$\Pi\subset\Gamma_{4,4}\otimes\R$. Let $\Sigma'=w^\perp\cap\Pi$ be a
space-like 3-plane and define the vector $B'$ by the fact that $\Pi$
is a spanned by $\Sigma'$ and $B'$ and $B'$ is perpendicular to any
vector in $\Sigma'$. We normalize $B'$ so that $B'.w=1$. We then
project $\Sigma'$ and $B'$ into $\Gamma_{3,3}\otimes\R$ to give the
space-like 3-plane $\Sigma$ and the vector $B$. We may write
\begin{equation}
  B' = \alpha w+w^*+B,
\end{equation}
for some real positive $\alpha$. 

Thus a point in $\Gr(\Gamma_{4,4})$ can be specified by a point in
$\Gr(\Gamma_{3,3})$, a vector in $\Gamma_{3,3}\otimes\R$ and a real
number $\alpha$. Different choices of $w\in\Gamma_{4,4}$ are
identified by the $\GO(\Gamma_{4,4})$ action. For a given $w$,
different choices of $w^*$ amount to shifts of $B$ by elements of
$\Gamma_{3,3}$.

We expect $\Gamma_{3,3}$ to be
the integral 2-cohomology of the K3 surface with two $E_8$
singularities, $\Sigma$ to represent
the Ricci-flat metric on $S_H$ in the usual way, $B$ to represent the
$B$-field and the volume of the K3 surface to be given by $2\alpha+
B^2$ \cite{me:lK3}.\footnote{This volume factor has errors in
earlier versions of \cite{me:lK3}.}

The 3-plane $\Sigma$ represents the directions in $H^2$ spanned by the
real and imaginary part of the holomorphic 2-form of $S_H$ and by the
K\"ahler form. Let $\Omega$ represent the two plane spanned by the
holomorphic 2-form and let $J$ be the K\"ahler form. Thus $\Sigma$ is
spanned by $\Omega$ and $J$. Rotating this decomposition of $\Sigma$
into $\Omega$ and $J$ amounts to changing the complex structure while
fixing the metric on $S_H$.

Let us fix a point in the moduli space $\Gr(\Gamma_{4,4})$ and fix a
choice of $w$ to give a geometrical interpretation.
Let $e$ be any vector in $\Gamma_{3,3}$ such that $e.e=-2$. Let us
choose $\Omega$ so that it is orthogonal to $e$. We are always free to
do this for a fixed $\Sigma$. Let us suppose that we are in the
generic situation where $e$ 
is then the only such vector orthogonal to $\Omega$. In this case $e$
(or $-e$) must represent a rational curve in $S_H$ (see for example 
\cite{W:K3inst} for this argument). Such a rational curve is a minimal
2-sphere are thus can give instantons. Thus all such $-2$ vectors (for
one choice of sign) give instantons.

To compute the area of such a curve we may simply take the dot product
of $e$ with $J$ using the normalization $J.J=2\alpha+B^2$. Thus we may
compute the associated area of all of the instantons without too much
difficulty in principle from this lattice picture of the moduli space.

If $S_h$ is elliptic with a section we may further decompose
$\Gamma_{3,3}=\Pic\oplus\Gamma^T_{2,2}$. Let $\Pic\cong\Gamma_{1,1}$
be spanned by light-like vectors $v$ and $v^*$. The complex structure
implied by demanding that $S_H$ be elliptic with section implies that
we choose $\Omega\subset\Gamma^T_{2,2}\otimes\R$ and 
$J\subset\Pic\otimes\R$. Let $\tilde J=\beta v+v^*$,
where $\tilde J$ is proportional to $J$. This fixes
\begin{equation}
  J = \sqrt{\frac{(2\alpha+B^2)\beta}2}\,v+
     \sqrt{\frac{2\alpha+B^2}{2\beta}}\,v^*.
\end{equation}
We are free to declare that the class $v$ represents the elliptic fibre and
$v^*-v$ represents the section. Thus the area of the fibre is
\begin{equation}
  J.v = \sqrt{\frac{2\alpha+B^2}{2\beta}},
\end{equation}
and the area of the section is
\begin{equation}
  J.(v^*-v) = \sqrt{\frac{(2\alpha+B^2)\beta}2}-
	\sqrt{\frac{2\alpha+B^2}{2\beta}}.
\end{equation}
So if we take the limit $\alpha\to\infty$ then both the fibre and the
section are infinite for generic $\beta$. Indeed for generic complex
structure we see that {\em all\/} 2-spheres in $S_H$ are infinite as
$J.J$ must be normalized to infinity.

To move back from this limit we may let $\alpha$ be finite. It would
be nice if we could still keep the area of the section infinite as
this would correspond to keeping the type IIA string weakly
coupled. We can see we may do this if we allow $\beta\to\infty$ while
keeping $\alpha$ finite (but large). {\bf This is the limit we use in
this paper.} 

Note that in this limit the area of the fibre is now zero! This should
not be seen as particularly nasty however as we don't expect the
elliptic fibre to generate any world-sheet instantons. What is more
interesting is that since $\alpha$ is finite, the normalization of $J$
is finite and thus for generic values of complex structure, the
transcendental 2-spheres in $S_H$ will have {\em finite and nonzero\/}
size. What we have is a rather funny-looking K3 surface which has an
enormous 2-sphere as a section, vanishingly small elliptic fibres and
transcendental 2-spheres which can have any size we wish. The volume
of the K3 surface is finite.

\iffigs
\begin{figure}
\begin{center}
  \epsfysize=7.5cm
  \leavevmode\epsfbox{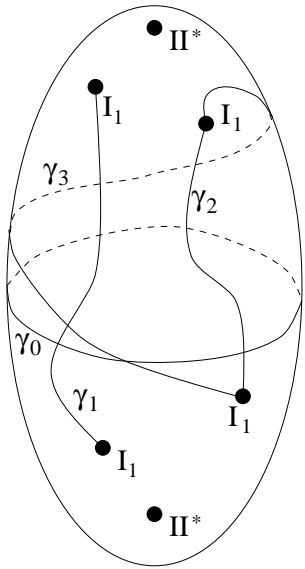}
\end{center}
  \caption{One-chains producing transcendental 2-cycles.}
  \label{fig:gs}
\end{figure}
\fi

It is therefore these transcendental 2-spheres, i.e. the length $-2$ vectors
in $\Gamma_{2,2}^T$,
which will be of most interest to us. These are the ones which give
nontrivial world-sheet instanton effects.

These transcendental 2-spheres may be visualized without too much
difficulty in terms described in \cite{me:MvF}. Figure~\ref{fig:gs}
depicts the spherical base of the fibration together with
the location of the degenerate fibres using Kodaira's notation.
Note the two $E_8$ singularities given by $\mathrm{II}^*$ fibres. 
Two transcendental 2-spheres may
be made by following the fibration along particular lines joining
$\mathrm{I}_1$ fibres. The idea is that a circle in the elliptic fibration
shrinks to zero size at the $\mathrm{I}_1$ ends and so this circle sweeps out a
2-sphere. The paths $\gamma_1$ and $\gamma_2$ in the figure show such
paths. The equatorial path $\gamma_0$ in the figure has no monodromy
and so we may make two transcendental tori by transporting either of
the two circles in the elliptic fibre along this path.

We may also may 2-spheres which are combinations of these classes. For
example path $\gamma_3$ gives a 2-sphere which is homologically
equivalent to the sum of the $\gamma_2$ sphere and one of the tori
from $\gamma_0$.

An important thing to note is that these 2-spheres never see the
$\mathrm{II}^*$ fibre and as such they miss anything interesting about the the
bundle over $S_H$. We pushed all the point-like instantons into
these $\mathrm{II}^*$ fibres. Thus {\em the bundle over the 2-spheres
we consider for worldsheet instantons is always trivial.}

There is an awkward point at this stage which we will evade in this
paper but which needs to understood more carefully if we want to
understand heterotic worldsheet instantons from type II
duality. Consider a transcendental 2-sphere whose class is given by
$e\in\Gamma_{2,2}^T$. We may then consider an infinity of other
2-spheres whose class is given by $e+nv$ for any $n\in\Z$. In the
limit $\beta\to\infty$ all of these 2-spheres will have exactly the
same area as the one given by $e$. It appears as though this fact makes
any instanton sum divergent. Clearly a more careful treatment of the
instanton sum in this limit is required. 


\section{Comments on the Type IIB Picture}   \label{s:IIB}

In this section we briefly note some peculiarities of the quantum
corrections to the type IIB picture of our model. This section is not
relevant to the following sections and may be skipped if the reader is
only interested in the hyperk\"ahler limits.

As discussed in the introduction, the type IIB string compactified on
$Y$ will suffer from both quantum corrections in $\alpha'$ and quantum
corrections in $\lambda$. It will therefore generically exhibit
worldsheet instanton corrections and spacetime instanton
corrections. What is perhaps surprising is that our picture in the
last section implies that it also exhibits ``mixed'' instanton
corrections depending on $\alpha'$ and $\lambda$ simultaneously.

To see this begin with the $\lambda\to0$ limit in which the type IIB
string has zero coupling and the section $\P^1$ of $S_H$ is infinitely
large. We have worldsheet instantons in the type IIB picture coming
from rational curves in $Y$. We also have worldsheet instanton
corrections in the 
heterotic picture coming from transcendental 2-spheres in $S_H$. If
$Y$ is rescaled to make it larger, the rational curves in $Y$ become
large suppressing the worldsheet instanton
corrections. In the heterotic picture, the corresponding deformation
makes the transcendental 2-spheres in $S_H$ larger resulting in a
smaller contribution from the worldsheet instantons.

Now let $\lambda$ acquire a small nonzero value for the type IIB
string. We know that the type IIB string will now pick up corrections
from spacetime instantons. In the heterotic picture the section of
$S_H$ now becomes finite and we may wrap worldsheet instantons around
this $\P^1$. Thus the spacetime instantons of the type IIB string
appear to map to heterotic worldsheet instantons wrapping the section
of $S_H$ while the worldsheet instantons of the type IIB string
correspond to  heterotic worldsheet instantons wrapping the
transcendental 2-spheres of $S_H$.

As explained by Witten \cite{W:K3inst} one should get heterotic
worldsheet instantons from {\em all\/} the 2-spheres in $S_H$
(assuming the bundle is suitably trivial). In
particular there are infinitely many 2-spheres which are homologically
a combination of the section of $S_H$ and the transcendental 2-spheres
discussed in the last section. What do heterotic instantons on these
``mixed'' 2-spheres correspond to in the type IIB string? Clearly they
give instanton corrections which depend both on $\alpha'$ and
$\lambda$. As such they are instantons that cannot decide whether they
should be considered ``worldsheet'' or ``spacetime'' instantons!
These mixed instantons must be taken into account beyond the quantum
corrections discussed in \cite{BBS:5b} for example if one is to obtain
an exact answer.


\section{The Hyperk\"ahler Limit}   \label{s:hypK}

One of the main sources of technical difficulties in dealing with the
hypermultiplet moduli space $\cM_H$ is that it is quaternionic
K\"ahler and very little is known about the general structure of
quaternionic K\"ahler spaces. We can be less ambitious in the rest of
this paper by only trying to ask about the rigid hyperk\"ahler limit
of properties of the moduli space. We should emphasize that up to this
point however our method should be perfectly good in principle for
determining some properties of the full quaternionic K\"ahler moduli space.

Taking the hyperk\"ahler limit is completely analogous to asking about
the rigid limit of special K\"ahler geometry. See \cite{Frd:SK,Craps:S} for
example for a discussion of the differences between rigid and nonrigid special
K\"ahler geometry.

Indeed in the very example we are looking at, the rigid limit of the
vector multiplet moduli space was studied in \cite{KKL:limit}. More
general analysis of this limit was explored in \cite{KKV:geng}.

In the language of the heterotic string compactified on a K3 surface
the hyperk\"ahler limit is understood as follows. We may remove
quantum corrections by taking the volume of the K3 surface to infinity
for a generic complex structure. Suppose instead we have a K3 surface
whose complex structure has been carefully chosen so that one minimal
2-sphere has a small volume. We may try to tune 
the complex structure so that the 2-sphere maintains its finite size
as we scale the overall volume of the K3 to infinity. What we should
be left with is a moduli space for this solitary 2-sphere sitting in a
non-compact space. We may similarly treat a collection of intersecting
2-spheres. The result is that we replace the moduli space of a compact
K3 surface by the moduli space of an ALE space.

Thus we need to zoom in on a part of the moduli space of complex
structures where a 2-sphere is of zero size as we take the stable
degeneration limit. This is very easy to do thanks to the work in
\cite{KKL:limit}. In particular we are very close to the analysis of
section 6 of \cite{KKL:limit}.

Let us consider a single 2-sphere shrinking down to an $A_1$
singularity for the first case. The stable degeneration is $y\to0$ as
discussed in section \ref{s:lim1}. We need to tune $x$ and $z$ so as
to hit the right part of $\Grm(\Gamma_{2,2}^T)$ to give a vanishing
2-sphere. We may do this by hitting a generic point on the discriminant
of (\ref{eq:jj1})
\begin{equation}
  (4z-1)(864^2z-(x-432)^2)=0.  \label{eq:A1}
\end{equation}
If we were talking about vector multiplets as in \cite{KKL:limit} this would
correspond to zooming in on an $\SU(2)$ Seiberg--Witten theory. 

Let $u$ be the complex variable specifying a generic slice through
the $x$-$z$-plane such that $u=0$ and $y=0$ gives a point on the
$\SU(2)$ locus (\ref{eq:A1}). Away from but close to $y=0$ there are
two solutions to the discriminant in the $u$-plane. To take the rigid
limit we take $y\to0$ while rescaling the $u$-plane so as to keep
these two solutions at fixed values, say $u=\pm\Lambda^2$.

The difference with the hypermultiplet moduli space as opposed to the
vector multiplet case is that we need to look at the R-R sector of the
type IIA moduli. These R-R fields live in the intermediate Jacobian
$H^3(X,\GU(1))$. As such we have a fibration over the above $u$-plane
whose fibre is an 8-dimensional torus.

The key point about the rigid limit of special K\"ahler geometry is
that the variation of Hodge structure of a \CY\ threefold is replaced
by the variation of a Hodge structure of a Riemann surface
\cite{Frd:SK,Craps:S}. As such, the intermediate Jacobian of the \CY\
threefold is replaced by the Jacobian of the corresponding Riemann
surface. This observation was effectively noted in \cite{GMV:Ht,OV:Din}.

Thus we don't need to do any work to say how the R-R fibres behave as
we move about the $u$-plane. The mini-variation of Hodge structure we
have zoomed in on by going to the rigid limit ignores the majority of
$H^3(X,\Z)$. We only care about the part which has nontrivial monodromy
around the $u$-plane. Thus we only need to consider two of the real
R-R fields. These will form an elliptic fibration over the $u$-plane
and of course this is exactly the elliptic fibration considered by
Seiberg and Witten in their original paper \cite{SW:I}.

We arrive at the conclusion that the interesting part of the moduli
space describing the acquiring of an $A_1$ singularity in the heterotic
string consists of a space with 4 real dimensions consisting of the
natural elliptic fibration over the Seiberg--Witten plane for an $\SU(2)$
gauge theory (with no flavours).

Given that the moduli space is hyperk\"ahler and thus Ricci-flat, this
describes the moduli space precisely except for one thing. We need
to specify the area of the elliptic fibre. This is the part of the
analysis that is quite difficult in our picture. To do this properly
we need to be careful about the rate at which we ``zoom in'' on the
moduli space to obtain the rigid limit while we take the type IIA
string to be weakly-coupled. Since analyzing the type IIA string at
nonzero coupling is rather difficult we will not attempt such an
analysis here. What is more, we already know the answer anyway thanks
to Seiberg and Witten \cite{SW:3d,W:hADE}. The result is that the
elliptic curve should be taken to have infinite area. In this case we
recover the Atiyah--Hitchin moduli space of 2 $\SU(2)$ monopoles
\cite{AH:mon}. This is also the same as the moduli space of $N=4$
$\SU(2)$ Yang--Mills in 3 dimensions.

It should now be immediately clear that we may use this same example
to probe the heterotic string near an $A_2$ singularity. The point in
$\Grm(\Gamma_{2,2}^T)$ which gives such a singularity is given by
$j_1=j_2=0$ and thus by $x=0$ and $z=\ff14$. This is exactly the point
studied in \cite{KKL:limit} to recover $\SU(3)$ Seiberg--Witten
theory. Thus we claim that the corresponding moduli space for
hypermultiplets is an abelian fibration over the $\SU(3)$
Seiberg--Witten theory where the abelian fibre is given by the
Jacobian of the Seiberg--Witten curve for $\SU(3)$ and the volume of
this fibre we take to be infinite again. This is the same as the
moduli space of $N=4$ $\SU(3)$ Yang--Mills in 3 dimensions by the
arguments in \cite{SW:3d}.


\section{Point-Like Instantons}   \label{s:pi}

We may modify this example to ``liberate'' one of the point-like $E_8$
instantons from one of the singularities in $S_H$. In other words we
need to take $X$ through an extremal transition to allow one more
deformation of complex structure. Let us call this required \CY\ space
$X_1$. Thus $h^{2,1}(X_1)=h^{2,1}(X)+1$.

\iffigs
\begin{figure}
\begin{center}
  \epsfysize=5cm\leavevmode\epsfbox{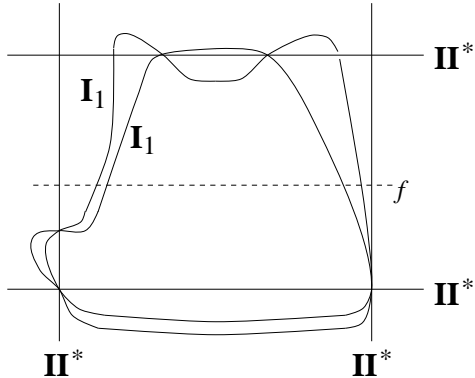}
\end{center}
  \caption{One point-like $E_8$ instanton liberated.}
  \label{fig:xi1}
\end{figure}
\fi

We may obtain $X_1$ by directly trying to engineer one $E_8$
point-like instanton at a generic point on $S_H$ while keeping
everything the same as the previous situation. From standard F-theory
lore \cite{AM:po,me:hyp} we need to replace the upper-right diagram in
figure~\ref{fig:squig} by that in figure~\ref{fig:xi1}. The collision
on the left side of the square represents to location of the
point-like instanton. This is explained in detail in \cite{AD:tang}.

Working backwards to the corresponding upper-left picture in
figure~\ref{fig:squig} it is not hard to show that $X_1$ is defined by
the equation
\begin{equation}
  f_{X_1}=f_X+a_8x_2^6x_3^7x_4^5,
\end{equation}
where $f_X$ is the defining equation of $X$ (written as a singular
form of $Y$) given by (\ref{eq:eqnX}).

Thus, the analogue of the geometric engineering story we did in the
last section is to consider what theory we would engineer in four
dimensions if we compactified the type IIA string on $Y_1$, the mirror
of $X_1$. Our
desired moduli space would then correspond to the hyperk\"ahler moduli
space given by reducing this four dimensional theory to one in three
dimensions.

To understand the geometry of $Y_1$ we have depicted a part of the
corresponding toric picture in figure~\ref{fig:tor1}. This diagram
shows a slice of the toric data representing $Y_1$ \`a la Batyrev
\cite{Bat:m}. Each vertex represents a divisor in $Y_1$ and
corresponds to a monomial in the mirror
$X_1$ \cite{AGM:mdmm}. We have labelled each vertex accordingly. The
vertex labelled  by $a_2$ represents the divisor given to first
approximation by $\P^1\times
C_0$ for some curve $C_0$. This is
obtained by blowing up the curve $C_0$ of $\Z_2$-fixed
points. The lines $C_1$ and $C_2$ in the figure represent two $\P^1$'s
with normal bundle $\O(-1,-1)$ which replace the generic $\P^1$ for a
particular point on $C_0$. Thus the divisor corresponding to $a_0$ is
a bundle over $C_0$ where every fibre is $\P^1$ except for one point
where this fibre splits into two $\P^1$'s. 

What we have described is precisely the geometric engineering picture
of an $\SU(2)\times\GU(1)$ gauge theory in four dimensions with a
hypermultiplet in the $\mathbf{2}$ of $\SU(2)$ charged with respect to
the $\GU(1)$ \cite{KKV:geng}. 

\iffigs
\begin{figure}
\begin{center}
  \epsfysize=5cm\leavevmode\epsfbox{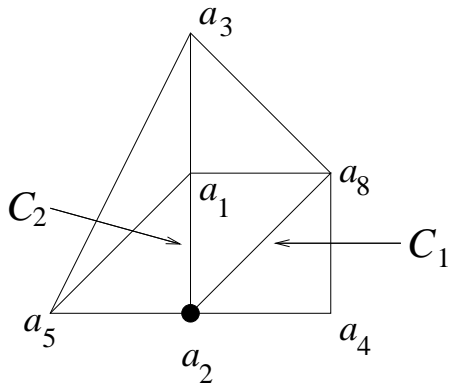}
\end{center}
  \caption{A Toric Picture for $Y_1$.}
  \label{fig:tor1}
\end{figure}
\fi

{\em Thus we arrive at the conclusion that the act of letting a free
point-like $E_8$ instanton ``interact'' with our $A_1$ singularity is
tantamount to adding a flavour to the Seiberg--Witten theory and a
gauging the associated $\GU(1)$ ``baryon number'' symmetry.}

Without much effort one can see that adding the term
$a_9x_2^6x_3^5x_4^7$ gives us another \CY\ space $X_2$ which
corresponds to {\em two\/} point-like instantons floating about the K3
surface $S_H$. $Y_2$ then has a ruled surface where one of the fibres
becomes a chain of three $\P^1$'s. This is the geometry of the picture
of an $\SU(2)\times\GU(1)^2$ gauge theory in four dimensions
with two charged $\mathbf{2}$'s \cite{KKV:geng}. 

Although we will not try to prove the general case, the result is
fairly self-evident leading to proposition \ref{p:d1} of the introduction.
Note in particular that once we have 4 or more point-like instantons
at an $A_1$ singularity, we obtain more gauge symmetry and so we have
a phase transition leading to more vector multiplets in the
four-dimensional theory. This is in perfect correspondence to the fact
that Seiberg--Witten theory for $\SU(2)\times\GU(1)^k$ has a Higgs
branch for $k\geq4$.

It is interesting to note that it was shown in \cite{AM:po} that $k\ge
2n$
point-like $E_8$ instantons coalescing at an $A_{n-1}$ singularity yields
a theory with a gauge group in the classical limit given
by\footnote{We assume there are no global properties of the
compactification leading to non-simply-connected gauge groups.}
\begin{equation}
\cG_{n,k} \cong 
  \SU(2)\oplus\SU(3)\oplus\ldots\oplus\SU(n-1)\oplus
     \SU(n)^{\oplus(k-2n+1)}\oplus\SU(n-1)\oplus\ldots\oplus\SU(2),
\label{eq:bigroup}
\end{equation}
with bifundamental hypermultiplets charged under each adjacent pair in
(\ref{eq:bigroup}), as well as extra fundamentals under the two $\SU(2)$
factors and under the first and last $\SU(n)$'s.
As such, once we include quantum corrections we expect the
moduli space of the {\em Coulomb\/} branch of our four-dimensional
theory to be given by the Seiberg--Witten theory of this group
$\cG_{n,k}$.

The astute reader will immediately notice that this yields precisely 
the theory in which in three dimensions is the 
Intriligator--Seiberg mirror of a theory with gauge symmetry
$\SU(n)\times\GU(1)^k$ \cite{IS:3dmir,HW:3d}.

This is explained as follows (see also \cite{SS:3dU} and particularly
\cite{HOV:mir}\footnote{Note that any local analysis along the lines of
\cite{HOV:mir} can often produce extraneous $\GU(1)$'s in the gauge
group.} for some
closely-related comments). Recall the following language for an
extremal transition. When in four dimensions the Coulomb phase is part
of the moduli space of $\cM_V$ and is thus special K\"ahler where the
Higgs phase is part of $\cM_H$ and is thus quaternionic K\"ahler. If
we dimensionally reduce this to three dimensions there 
is no intrinsic distinction between vector multiplets and
hypermultiplets --- they both give a quaternionic K\"ahler moduli space. Even
including quantum corrections these multiplets do not mix however and
we may still logically label them by their four dimensional origin.

We have argued that the moduli space of $k$ instantons on an $A_{n-1}$
singularity is given by the Coulomb branch of an
$\SU(n)\times\GU(1)^k$ theory in three dimensions. This has a phase
transition to a Higgs branch. The results of \cite{AM:po} tell us that
this Higgs branch is actually given by the Coulomb branch of a theory
with gauge group $\cG_{n,k}$ in three dimensions. Thus the Higgs
branch of $\SU(n)\times\GU(1)^k$ is isomorphic to the Coulomb branch
of $\cG_{n,k}$. This is a statement of Intriligator--Seiberg mirror
symmetry. Here we have deduced it from the mirror symmetry of an $X$-type
manifold whose stable degeneration represents the heterotic string
picture and the mirror $Y$-type which would be used to geometrically
engineer the corresponding moduli space.

Note that we could not have used Intriligator--Seiberg mirror symmetry
to deduce the examples we discussed for small $k$. Here the theory has
no phase transition.

Given the plethora of heterotic string pictures which are understood
in F-theory \cite{MV:F2,AM:po,AM:frac} one should be able to extend the
analysis of this paper to many more examples.

\section*{Acknowledgements}

It is a pleasure to thank S.~Katz, D.~Morrison, S.~Sethi, E.~Sharpe and
E.~Witten for useful discussions. P.S.A.~is supported in part by the
Alfred P. Sloan Foundation.  M.R.P. thanks the Aspen Center for
Physics for a stimulating workshop where some of this work was completed.


\end{document}